\documentclass[preprint]{aastex}

\begin{document}

\title {Absolute Properties of the Pulsating Post-mass Transfer Eclipsing Binary OO Draconis}
\author{Jae Woo Lee$^{1,2}$, Kyeongsoo Hong$^{1,3}$, Jae-Rim Koo$^{1,4}$, and Jang-Ho Park$^{1,5}$ }
\affil{$^1$Korea Astronomy and Space Science Institute, Daejeon 34055, Korea}
\affil{$^2$Astronomy and Space Science Major, Korea University of Science and Technology, Daejeon 34113, Korea}
\affil{$^3$Institute for Astrophysics, Chungbuk National University, Cheongju 28644, Korea}
\affil{$^4$Department of Astronomy and Space Science, Chungnam National University, Daejeon 34134, Korea}
\affil{$^5$Department of Astronomy and Space Science, Chungbuk National University, Cheongju 28644, Korea}
\email{jwlee@kasi.re.kr, kshong@kasi.re.kr, koojr@kasi.re.kr, pooh107162@kasi.re.kr}

\begin{abstract}
OO Dra is a short-period Algol system with a $\delta$ Sct-like pulsator. We obtained time-series spectra between 2016 February 
and May to derive the fundamental parameters of the binary star and to study its evolutionary scenario. The radial velocity (RV) 
curves for both components were presented, and the effective temperature of the hotter and more massive primary was determined 
to be $T_{\rm eff,1}$ = 8260 $\pm$ 210 K by comparing the disentangling spectrum and the Kurucz models. Our RV measurements 
were solved with the $BV$ light curves of Zhang et al. (2014) using the Wilson-Devinney binary code. The absolute dimensions 
of each component are determined as follows: $M_1$ = 2.03 $\pm$ 0.06 M$_\odot$, $M_2$ = 0.19 $\pm$ 0.01 M$_\odot$, 
$R_1$ = 2.08 $\pm$ 0.03 R$_\odot$, $R_2$ = 1.20 $\pm$ 0.02 R$_\odot$, $L_1$ = 18 $\pm$ 2 L$_\odot$, and 
$L_2$ = 2.0 $\pm$ 0.2 L$_\odot$. Comparison with stellar evolution models indicated that the primary star resides inside 
the $\delta$ Sct instability strip on the main sequence, while the cool secondary component is noticeably overluminous and 
oversized. We demonstrated that OO Dra is an oscillating post-mass transfer R CMa-type binary; the originally more massive star 
became the low-mass secondary component through mass loss caused by stellar wind and mass transfer, and the gainer became 
the pulsating primary as the result of mass accretion. The R CMa stars, such as OO Dra, are thought to have formed by 
non-conservative binary evolution and ultimately to evolve into EL CVn stars.
\end{abstract}

\keywords{binaries: eclipsing --- stars: fundamental parameters --- stars: individual (OO Dra) --- stars: oscillations (including pulsations) --- techniques: spectroscopic}{}

\section{INTRODUCTION}

Classical Algol-type eclipsing stars are generally semi-detached binary systems in which a lobe-filling secondary with 
a spectral type of F-K is transferring its mass to an early-type main-sequence primary component. Because the secondary components 
are magnetically active, the binary stars may lose both mass and angular momentum from the systems via magnetic stellar winds and, 
hence, the mass transfer between the component stars should be non-conservative (Sarna 1993; Erdem \& \"Ozt\"urk 2014). 
The Algol systems can be considered a good astrophysical laboratory for studying mass-transfer processes and binary evolution. 
For this purpose, it is necessary to precisely measure the physical parameters of each component by time-series photometry and 
spectroscopy. Double-lined eclipsing binaries (EBs) are valued because they allow for a direct and accurate determination of 
mass, radius, and luminosity (Hilditch 2001). However, because the light contribution from the cool secondary in an Algol star 
is only a few percent of the total brightness of the eclipsing system, it is not easy to separate the two components from 
the observed spectra. Nonetheless, the spectrum of the faint component can be reconstructed (e.g., Matson et al. 2015).

Pulsating stars provide significant insight into the stellar interior structure based on an analysis of their oscillation 
frequencies. Pulsating EBs that show both eclipses and pulsations are very interesting subjects for investigating the structure 
and evolution of stars through their binary properties and asteroseismology (e.g., Lee 2016; Lee et al. 2016). There are many EBs 
with various types of pulsating components, such as $\gamma$ Dor (\c Cakirli \& Ibano\v{g}lu 2016), red giant (Gaulme et al. 2013), 
and subdwarf B (Schaffenroth et al. 2015) pulsators. Over 90 of them have been known to be semi-detached EBs with 
$\delta$ Sct-type components (Mkrtichian et al. 2004; Zhou 2010). The oscillating primary components show pulsation behaviors 
analogous to those of single $\delta$ Sct pulsators, but they may have a different evolutionary history because they evolved 
with tidal interaction and mass accretion between both components. Thus, pulsations in EBs may be affected by binary effects. 
Recently, Kahraman Ali\c cavu\c s et al. (2017) examined the physical parameters of pulsating EBs and their possible relationships, 
and compared them with those of single $\delta$ Sct variables. They showed that the $\delta$ Sct components in binary stars 
pulsate at lower amplitudes and for shorter periods than single pulsators. However, the physical properties of pulsating EBs 
still have not been thoroughly studied because of the absence of observed spectra and radial velocities (RVs) in many cases 
(Liakos \& Niarchos 2017; Kahraman Ali\c cavu\c s et al. 2017). 

To obtain high-resolution spectra of pulsating EBs and to study their absolute properties, we chose the Algol system OO Dra 
(TYC 4550-1408-1, 2MASS J11400140+7509215, NSVS 879987; $V\rm_T$ = $+$11.20, $(B-V)\rm_T$ = $+$0.27). The program target was 
discovered to be an EB of about 1.238 d period by Biyalieva \& Khruslov (2007). Dimitrov et al. (2008) presented 
their partial $BVR$ light curves and single-lined RV data. A pulsation frequency of $\sim$37 cycle d$^{-1}$ was detected by 
a preliminary analysis of the outside-eclipse photometric observations. From the observed spectra, they determined 
the surface gravity, effective temperature, and projected rotational velocity of the primary component to be $\log g$ = 4.0, 
$T_{\rm eff,1}$ = 8500 K, and $v_1\sin$$i$ = $\sim$60 km s$^{-1}$, respectively. Zhang et al. (2014) presented 
their $BV$ photometric observations, and reported that the eclipsing binary may be a detached system with the cool secondary 
almost (97.5 \%) filling its limiting lobe. From eclipse-subtracted light residuals in both bands, they detected two frequencies 
of 41.87 cycle d$^{-1}$ and 34.75 cycle d$^{-1}$ and confirmed the $\delta$ Sct-like pulsation nature of the EB system. 

This article is the fourth contribution in a series of detailed studies, based on time-series spectroscopic observations of 
pulsating EBs (Hong et al. 2015, 2017; Koo et al. 2016). Here, we present OO Dra as a pulsating post-mass transfer R CMa star 
evolving from a semi-detached Algol into an EL CVn-type binary, a new class of EBs defined by Maxted et al. (2014a).

\section{OBSERVATIONS AND DATA ANALYSIS}

Spectroscopic observations of OO Dra were made using a fiber-fed high-resolution echelle spectrograph BOES (Kim et al. 2007) 
attached to the 1.8 m reflector at Bohyunsan Optical Astronomy Observatory (BOAO) in Korea. A total of 39 spectra were obtained 
during six nights between February and May 2016. The exposure time was set to 1800 s, considering the brightness and orbital period 
of the program target. The BOES has five optical fibers covering the wavelength range of 3600$-$10,200 $\rm \AA$ (Kim et al. 2007; 
Lee et al. 2008). We selected the largest fiber with a resolving power of $R$ = 30,000. Bias, Tungsten-Halogen lamp, and 
Th-Ar Arc images were also taken for pre-processing and wavelength calibration. The reduction process for the observed spectra 
was the same as that used by Hong et al. (2015). The signal-to-noise (S/N) ratio at around 5550 $\rm \AA$ was typically about 30.

The cross-correlation technique (CCT) is commonly used for RV determination in spectroscopy. However, the secondary component 
of OO Dra contributes $\la 10$ \% to the total luminosity of the binary system (Zhang et al. 2014), and the S/N ratios of 
our spectra were not sufficiently high; thus, it was difficult to apply the CCT to the observed spectra of this system. 
Dimitrov et al. (2008) also measured the RVs of the primary star, but were not able to detect the signal of the companion. 
Following a similar approach to that of Hong et al. (2015), we searched for absorption lines in our spectra that could be used 
to plainly identify the two component stars, and found an isolated double line of Fe I $\lambda$4957.61. Figure 1 displays 
the trailed spectra in the Fe I $\lambda$4957.61 region and a sample of spectral lines observed near an orbital phase of 0.75. 
The absorption lines were fitted by two Gaussian functions using the IRAF $splot$ task with a deblending option. 
The resultant RV measurements and their errors are given in Table 1. Our double-lined RVs are shown in Figure 2, together with 
the single-lined measures of Dimitrov et al. (2008) for comparison. 

The spectroscopic elements of OO Dra were obtained assuming a circular orbit ($e=0$). We applied a least-squares fit separately 
to the primary and secondary RV curves, using a sine wave of $V = \gamma + K \sin[2 \pi (t-T_{\rm 0})/ P]$. Here, $\gamma$ and 
$K$ are the systemic velocity and the semi-amplitude of each RV curve, respectively, $P$ and $T_{\rm 0}$ are the period and 
reference epoch of the orbital ephemeris, respectively, and $t$ is the observation time of each data point. The orbital period 
was fixed to be $P$ = 1.238378 d based on the eclipse timing analysis of Zhang et al. (2014). The sine-curve solutions are 
summarized in Table 2, where the systemic velocities determined from both components agree within their uncertainties. 
The velocity semi-amplitudes were combined to obtain a spectroscopic mass ratio of $q = 0.092\pm0.010$. The sine curves calculated 
with the orbital parameters of Table 2 are plotted in Figures 1 and 2. The rms residuals for each fit were found to be 3.3 km s$^{-1}$ 
and 7.8 km s$^{-1}$, respectively, for the primary and secondary star.

As stated above, because the S/N ratio and the light contribution of the cool secondary are low, it is not possible to derive 
reliable atmosphere parameters of the faint star. For measuring the atmospheric parameters of the more massive primary component, 
we obtained a high S/N disentangling spectrum using the FDBinary code (Iliji\'c et al. 2004)\footnote{\url{http://sail.zpf.fer.hr/fdbinary/}}. 
In the reconstructed spectrum, we chose four spectral regions that are temperature indicators for A0$-$F0 dwarf stars according 
to the \textit{Digital Spectral Classification Atlas} of R. O. Gray. A total of 17,850 synthetic spectra with intervals of 
7500 $\le$ $T_{\rm eff,1}$ $\le$ 11,000 K (in steps of 10 K) and 50 $\le$ $v_1\sin$$i$ $\le$ 100 km s$^{-1}$ 
(in steps of 1 km s$^{-1}$) were constructed by interpolating the ATLAS9 model atmospheres (Kurucz 1993). For the synthetic 
spectra, we assumed a solar metallicity of $[$Fe/H$]=0$ and a microturbulent velocity of 2.0 km s$^{-1}$. The surface gravity of 
the pulsating primary star was set to be log $g_1 = 4.1$ obtained from our binary solutions presented in the following section. 
However, it is possible that the real metallicity could be different from the solar value.

We searched for a global minimum for both effective temperature ($T_{\rm eff,1}$) and projected rotational velocity ($v_1\sin$$i$), 
calculating the $\chi^2$ statistics between the synthetic and disentangling spectra. The process was very similar to that 
used by Hong et al. (2017), and the search results are presented in Figure 3. We found the best-fitting parameters of 
$T_{\rm eff,1}=8260 \pm 210$ K and $v_1\sin$$i=72\pm5$ km s$^{-1}$. Their errors were estimated as the 1$\sigma$-values of 
the optimal values derived in each spectral region. Figure 4 represents the disentangling spectrum from the FDBinary code, 
together with three synthetic spectra of 8050 K, 8260 K, and 8470 K.

\section{BINARY MODELING AND ABSOLUTE DIMENSIONS}

In their light-curve modeling for OO Dra, Zhang et al. (2014) used the $q$-search method for a series of models with 
varying mass ratio and suggested two possible photometric solutions: a detached binary with $q=0.097$ and 
a semi-detached binary (the secondary component filling its limiting lobe completely) with $q=0.075$. The mass ratio is 
thus one of the most important parameters in studying the absolute properties of the binary system. To determine 
a unique solution for OO Dra, our double-lined RV data were analyzed with the $BV$ light curves of Zhang et al. (2014), 
considering proximity and eclipse effects. We used the 2007 version of the Wilson-Devinney binary code 
(Wilson \& Devinney 1971; van Hamme \& Wilson 2007; hereafter W-D) and a weighting scheme almost identical to that used 
for the eclipsing binaries V407 Peg (Lee et al. 2014) and DK Cyg (Lee et al. 2015). Table 3 lists the OO Dra curves and 
their standard deviations ($\sigma$) used in our synthesis.

The surface temperature of the hotter and more massive primary component was given as $T_{1}$ = 8260 K from our spectral 
analysis. The gravity-darkening exponents were assumed to be $g_1$=1.0 and $g_2$=0.32 (von Zeipel 1924; Lucy 1967) and 
the bolometric albedos $A_1$=1.0 and $A_2$=0.5 (Rucinski 1969a,b) because the primary and secondary components should 
have radiative and convective atmospheres, respectively. The logarithmic monochromatic ($x$, $y$) and bolometric ($X$, $Y$) 
limb-darkening coefficients were interpolated from van Hamme (1993). Furthermore, a synchronous rotation ($F_{1,2}$ = 1.0) 
for the component stars was adopted, and the detailed reflection effect was used throughout the analyses. Subscripts 1 
and 2 denote the primary and secondary components being eclipsed at Min I and Min II, respectively.

In principle, RV and light curves can be simultaneously modeled, but it is difficult to give reasonably different weights to 
them. To address this, our binary modeling was performed in two stages. In the first stage, we analyzed the $BV$ light curves 
of Zhang et al. (2014) with the spectroscopic elements ($\gamma$, $a$, and $q$) of Table 2 obtained by fitting a sinusoid to 
each RV curve. In the second, our double-lined RV measurements were solved using the photometric parameters yielded in 
the first stage. This procedure was iterated until both data sets were satisfied together. The final results are presented in 
columns (2)--(3) of Table 4. The synthetic RV curves are given as solid curves in Figure 2, while the synthetic light curves 
are displayed in Figure 5. In all procedures, we considered the orbital eccentricity as an additional free parameter, but 
its value remained zero, which implies that OO Dra is in a circular orbit. 

Our consistent light and RV solution indicates that OO Dra is in a detached configuration with a low mass ratio of 
$q$ = 0.092 and fill-out factors of $F_1$ = 61 \% and $F_2$ = 99 \%, where $F_{1,2}$ = $\Omega_{1,2}$/$\Omega_{\rm in}$. 
The absolute dimensions of the pulsating EB were calculated from the light and velocity parameters, and the results are given 
in Table 5. Luminosities ($L$) and bolometric magnitudes ($M_{\rm bol}$) were derived by using $T_{\rm eff}$$_\odot$ = 5780 K 
and $M_{\rm bol}$$_\odot$ = +4.73 for solar values. Bolometric corrections (BCs) were taken from the expression between 
$\log T_{\rm eff}$ and BC (Torres 2010). Using an apparent visual magnitude of $V$ = +11.079 $\pm$ 0.038 (Henden et al. 2016) 
and the interstellar reddening of $A_{\rm V}$ = 0.210 (Schlafly \& Finkbeiner 2011), we determined the distance to 
our program target to be 760 $\pm$ 42 pc. This value is consistent with 649 $\pm$ 122 pc computed from Gaia DR1 
(1.54 $\pm$ 0.29 mas; Gaia Collaboration et al. 2016) within their errors. 

We calculated the rotational velocities of both components to be $v_{\rm 1,sync}$ = 84.9 $\pm$ 1.1 km s$^{-1}$ and 
$v_{\rm 2,sync}$ = 49.0 $\pm$ 0.7 km s$^{-1}$, respectively, in the hypothesis of synchronous rotation. 
Our measured rotational velocity of $v_1$$\sin$$i$ = 72 $\pm$ 5 km s$^{-1}$ indicates that the primary star is spinning slower 
than the synchronized value. Therefore, we set the ratio of the axial rotation rate to the mean orbital rate to be $F_1$ = 0.85 
and reanalyzed the light and RV curves in Table 3. The new solutions from this value are in good agreement with those from 
$F_1$ = 1.0, and there are no significant differences in $\sum W(O-C)^2$ (the weighted sum of the squared residuals) between 
them. We can know that the $F_1$ value does not affect the astrophysical parameters presented in this article. On the other hand, 
the synchronization ($\log t_{\rm sync}=5.92\pm0.09$) and circularization ($\log t_{\rm sync}=5.30\pm0.03$) timescales of 
OO Dra may be shorter than its age, considering the evolutionary status of the R CMa-type Algols discussed in Section 4. 
The EB system is then expected to be in a synchronized rotation. Figures 2 and 5 show that our binary modeling with both 
a circular orbit and a synchronous rotation describes the RV and light curves quite well.

\section{DISCUSSION AND CONCLUSIONS}

In this article, we presented time-series spectroscopy of the pulsating Algol system OO Dra, which was conducted using 
the high-resolution BOES. From the observed spectra, we obtained a new RV curve for the primary star and the first RV curve 
for the faint secondary, with better coverage and precision than the single-lined RV1 data of Dimitrov et al. (2008). 
The analysis of the disentangling spectrum for the primary component gave us a temperature of $T_{\rm eff,1}=8260 \pm 210$ K 
and a projected rotational velocity of $v_1\sin$$i=72\pm5$ km s$^{-1}$. Using these and the orbital elements of Table 2, 
our double-lined RV curves were solved with the $BV$ photometric data of Zhang et al. (2014). The binary modeling indicates 
that our program target is a short-period detached system with a mass ratio of $q$ = 0.092 and an orbital inclination of 
$i$ = 85$^\circ$.3 and that the secondary's temperature is $T_1$ = 6268 K. The RV and light parameters allowed us to 
calculate the absolute dimensions for each component, and the results are the following: $M_1 = 2.03 \pm 0.06$ M$_\odot$, 
$M_2 = 0.19 \pm 0.01$ M$_\odot$, $R_1 = 2.08 \pm 0.03$ M$_\odot$, $R_2 = 1.20 \pm 0.02$ M$_\odot$, $L_1 = 18 \pm 2$ L$_\odot$, 
and $L_2 = 2.0 \pm 0.2$ L$_\odot$. 

From a frequency analysis of the eclipse-subtracted $BV$ light residuals, Zhang et al. (2014) detected two pulsation frequencies 
of $f_1$ = 41.87 cycle d$^{-1}$ and $f_2$ = 34.75 cycle d$^{-1}$, but the periodicity of $f_3$ $\approx$ 37 cycle d$^{-1}$ shown 
in the photometric data of Dimitrov et al. (2008) was below the significant limit because of the high noise. They suggested that 
OO Dra is an EB containing a $\delta$ Sct-type pulsator. Usually, $\delta$ Sct stars pulsate in low-order press ($p$) modes with 
short periods of 0.02$-$0.25 d and pulsation constants of $Q <$ 0.04 d (Breger 2000). Because the absolute dimensions of 
the primary component had been accurately measured, we applied them to the relation of 
$\log Q_i = -\log f_i + 0.5 \log g + 0.1M_{\rm bol} + \log T_{\rm eff} - 6.456$ (Petersen \& J\o rgensen 1972). Pulsation 
constants for the three frequencies ($f_{1,2,3}$) was determined to be $Q_1$ = 0.0113 d, $Q_2$ = 0.0136 d, and $Q_3$ = 0.0128 d. 
We compared the $Q$ values with the theoretical models with 2.0 $M_\odot$ (Fitch 1981). $f_1$ and $f_3$ could be identified 
as the sixth ($6H$) and fifth ($5H$) overtone radial modes, respectively, while $f_2$ as the fifth ($p_5$) non-radial mode of 
$\ell$ = 1. As shown in the HR diagram of Figure 6, the primary star resides in the $\delta$ Sct instability region. Further, 
the ratio, $P_{\rm pul}/P_{\rm orb}$ = 0.0193, of the pulsational to orbital periods is within the upper bound of 
$P_{\rm pul}/P_{\rm orb}$ = 0.09$\pm$0.02 for $\delta$ Sct pulsators in binaries proposed by Zhang et al. (2013). All of 
these results demonstrate that the oscillating primary component with a spectral type of A4V is a $\delta$ Sct-type variable. 
Although the $\delta$ Sct component in the binary exhibits practically the same pulsational characteristics ($P_{\rm pul}$ and $Q$) 
as normal $\delta$ stars, its pulsations might be affected by tidal and rotational distortion, and the mass transfer between 
the component stars could at least be partly responsible for the $\delta$ Sct frequencies. 

The physical properties of OO Dra are very similar to those of the pulsating EBs KIC 8262223 (Guo et al. 2017) and KIC 10661783 
(Lehmann et al. 2013), which are detached binary stars with characteristics (low $q$, $M_2$, and $P$ combinations) of R CMa 
(Budding \& Butland 2011) among Algols. At present, there have only been five R CMa-like binaries with accurate fundamental 
parameters from both time-series photometry and spectroscopy, including AS Eri (van Hamme \& Wilson 1984; Ibano\v{g}lu et al. 2006) 
and OGLEGC 228 (Kaluzny et al. 2007). Three non-{\it Kepler} targets (R CMa, AS Eri, OGLEGC 228) of these binaries are 
semi-detached systems with the secondary filling its critical Roche lobe, and four eclipsing binaries except for OGLEGC 228 
displayed $\delta$ Sct-type pulsations. We investigated the evolutionary state of OO Dra in terms of mass-radius, 
mass-luminosity, and HR diagrams. In Figure 6, the positions of the binary components are displayed as star symbols, together 
with well-studied classical Algols (circles; Ibano\v{g}lu et al. 2006). Here the lower and upper triangles denote detached and 
semi-detached R CMa-type Algols, respectively. As shown in the figure, both components of OO Dra conform to the general trend 
for R CMa stars. The primary component is located inside the main-sequence band in all diagrams, 
while the less massive secondary component is remarkably overluminous and oversized compared to dwarf stars of the same mass 
in the left two diagrams (M-R and M-L) but on the zero-age main sequence (ZAMS) in the HR diagram. These indicate that 
the initially more massive star became the current low-mass secondary component via non-conservative binary evolution, and 
the gainer became the present pulsating primary residing within the $\delta$ Sct instability region as the result of 
mass accretion (Chen et al. 2017; Hong et al. 2017). As in the case of the detached R CMa star KIC 8262223 (Guo et al. 2017), 
the secondary component of OO Dra stopped its mass transfer and, at present, is contracting. This would have caused 
the eclipsing system to evolve into the current detached state. 

In the HR diagram of Figure 6, we show the evolutionary track for helium white dwarfs (He WDs) with masses of 0.179 $M_\odot$ 
and 0.195 $M_\odot$ given by Driebe et al. (1998). Also, we plotted three EL CVn-type binaries of WASP J0247-25 
(Maxted et al. 2013), WASP J1628+10 (Maxted et al. 2014b), and KIC 9164561 (Zhang et al. 2016), which comprise a normal A(F)-type 
main-sequence star and a proto-He WD of $\sim$0.2$M_\odot$ (Maxted et al. 2014a; Chen et al. 2017). As shown in the figure, 
the secondary component (OO Dra B) of our program target with a mass of 0.187 $M_\odot$ is a good match to the models for 
the formation of He WDs. We think that OO Dra B is a precursor of a very low-mass He WD evolving to higher $T_{\rm eff}$ 
at almost constant $L$. Recently, Chen et al. (2017) investigated the formation of EL CVn binaries and showed that 
low-mass He WDs could be caused by non-conservative evolution through a stable mass transfer channel, but not 
rapid common-envelope evolution. We can consider the evolutionary state of OO Dra in their diagrams (Figures 6, 8, 10, and 11). 
The physical parameters of OO Dra B in Tables 4$-$5 match well with their $T_{\rm eff}-R/a$ and $M_{\rm WD}-P$ relations, which 
indicate that the secondary component has evolved from stable mass transfer and has a degenerate core. We estimated 
the envelope mass ($M_{\rm env}$) of OO Dra B and its lifetime ($t$) in the nearly constant-$L$ stage to be $\sim$0.02 $M_\odot$ 
and $\sim$6 $\times$ 10$^{8}$ yr, respectively, using the $M_{\rm WD}-M_{\rm env}$ and $M_{\rm WD}-t$ relations. The R CMa stars, 
including OO Dra, are thought to have formed as the result of non-conservative mass transfer and ultimately to evolve into 
EL CVn stars. However, because only six binaries have been identified as R CMa stars with reliable physical parameters, 
additional discoveries and follow-up observations will help clarify our understanding of the formation and characteristics of 
very low-mass WDs.

\acknowledgments{ }
The authors wish to thank the staff of BOAO for assistance during our observations. We also thank Dr. Xiaobin Zhang for 
sending us photometric data on OO Dra. This research has made use of the Simbad database maintained at CDS, Strasbourg, 
France, and was supported by the KASI grant 2017-1-830-03. The work by K. Hong and J.-R. Koo was supported by 
the grant numbers 2017R1A4A1015178 and 2017R1A6A3A01002871 of the National Research Foundation (NRF) of Korea.

\clearpage
\begin{figure}
\includegraphics[]{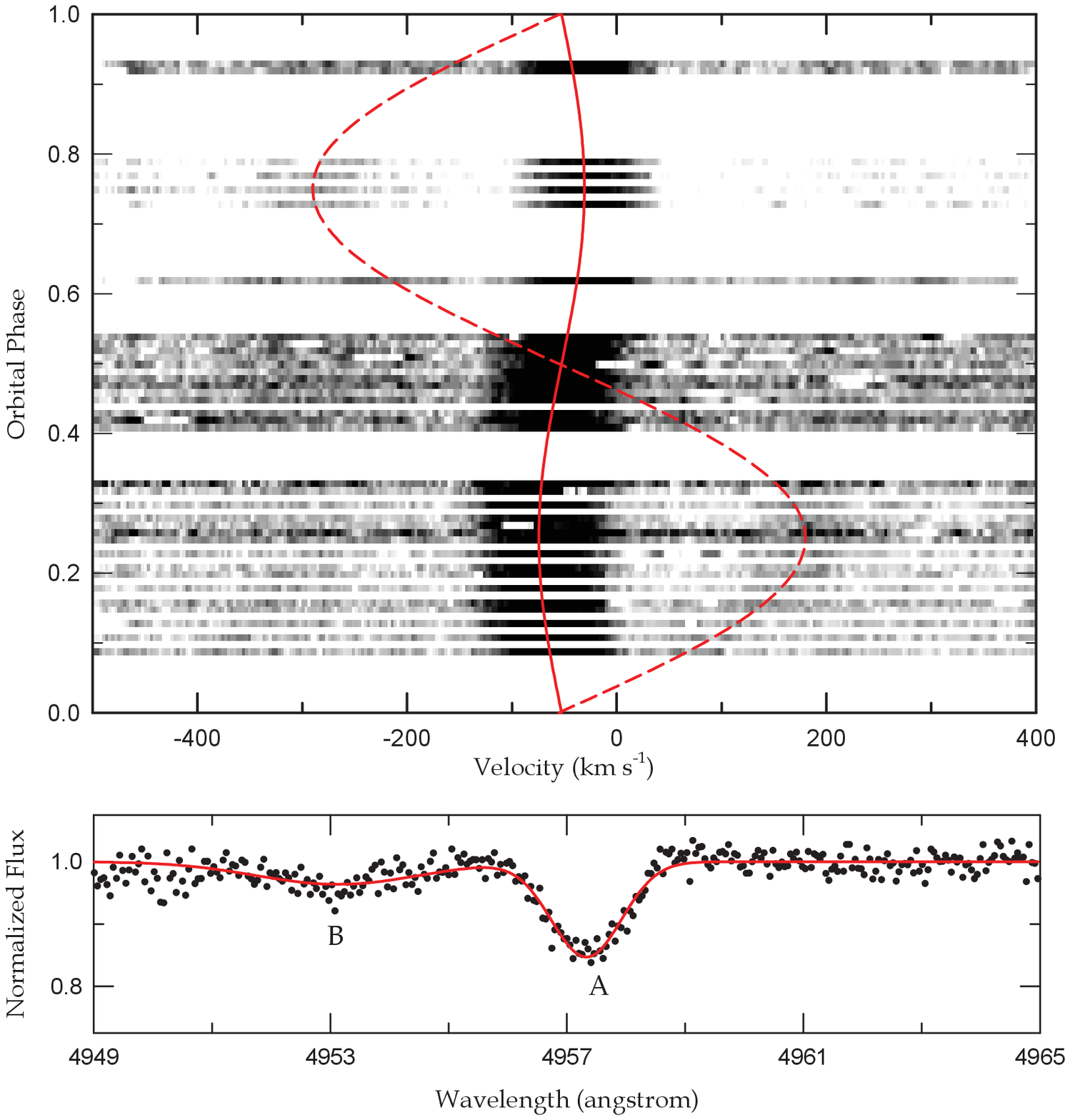}
\caption{The upper panel displays the trailed spectra of OO Dra in the Fe I $\lambda 4957.61$ region. The solid and 
dashed lines track the orbital motions of the primary (A) and secondary (B) components, respectively. In the lower panel, 
the circle and line represent the observed spectrum and its Gaussian fitting at an orbital phase of 0.75, respectively. }
\label{Fig1}
\end{figure}

\begin{figure}
\includegraphics[]{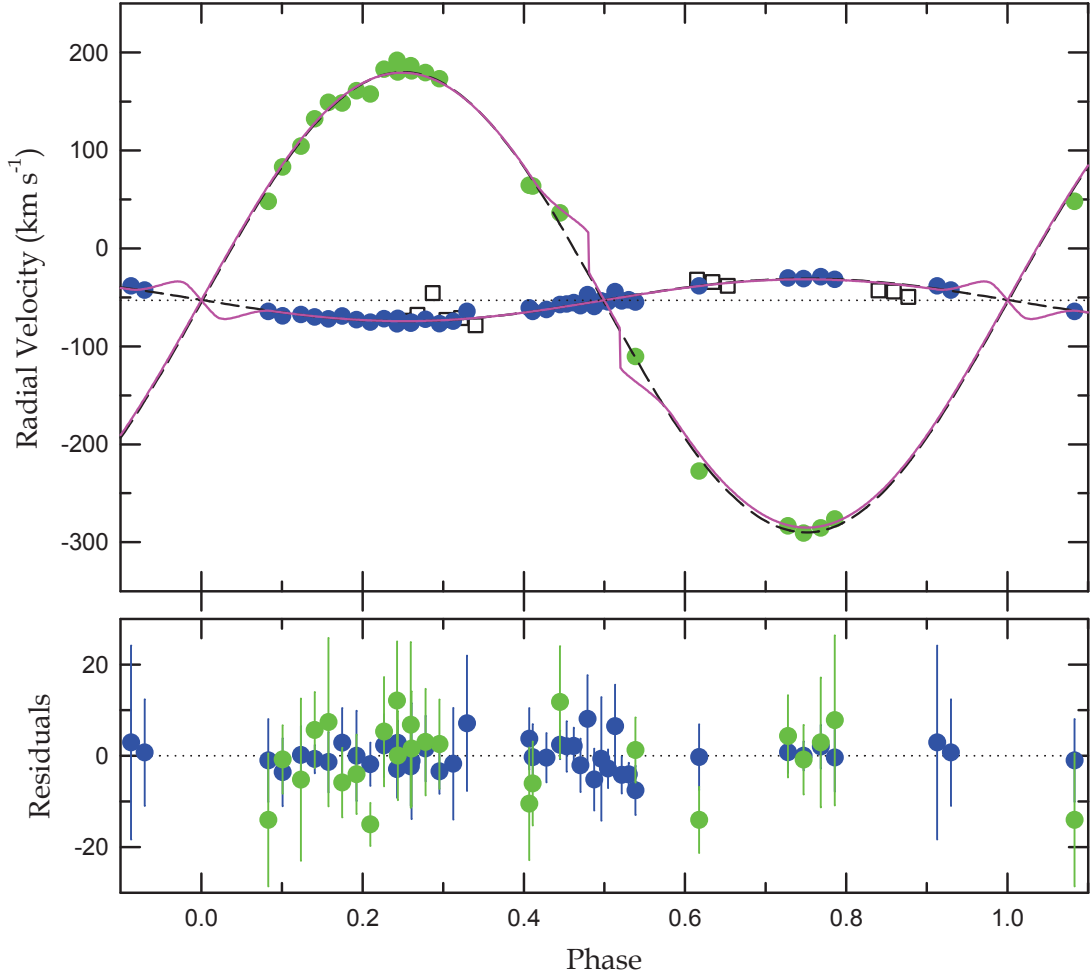}
\caption{Radial-velocity curves of OO Dra with fitted models. The squares and circles are the RV measures of Dimitrov et al. 
(2008) and ours, respectively. In the upper panel, the dashed curves were obtained by least-fitting separately a sine wave to 
our primary and secondary RVs, while the solid curves denote the results from a consistent light and RV curve analysis with 
the W-D code. The dotted line represents the system velocity of $-$52.8 km s$^{-1}$. The lower panel shows the residuals between 
our observations and sine-curve fits, where each vertical line is an error bar for each RV measurement. }
\label{Fig2}
\end{figure}

\begin{figure}
\includegraphics[scale=0.9]{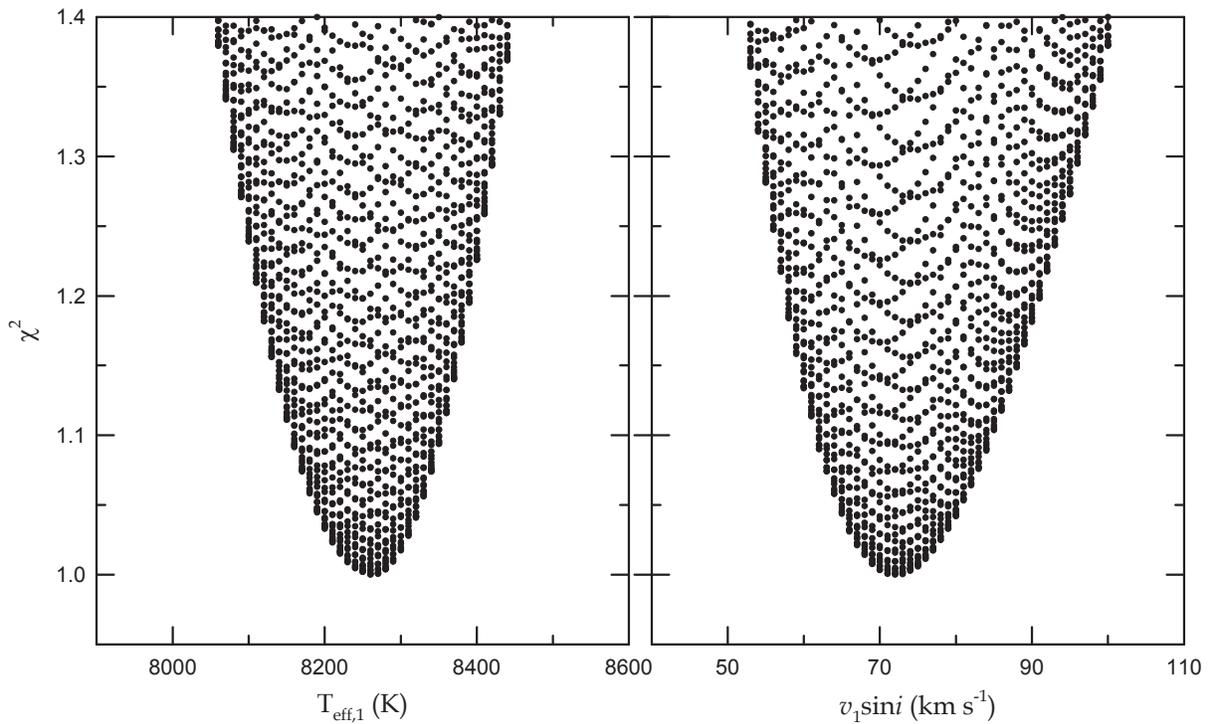}
\caption{$\chi^2$ diagrams of the effective temperature (left) and the projected rotational velocity (right) of the primary star. 
The surface gravity and metallicity are fixed as $\log$ $g$ = 4.1 and [Fe/H] = 0.0, respectively. }
\label{Fig3}
\end{figure}

\begin{figure}
\includegraphics[scale=0.8]{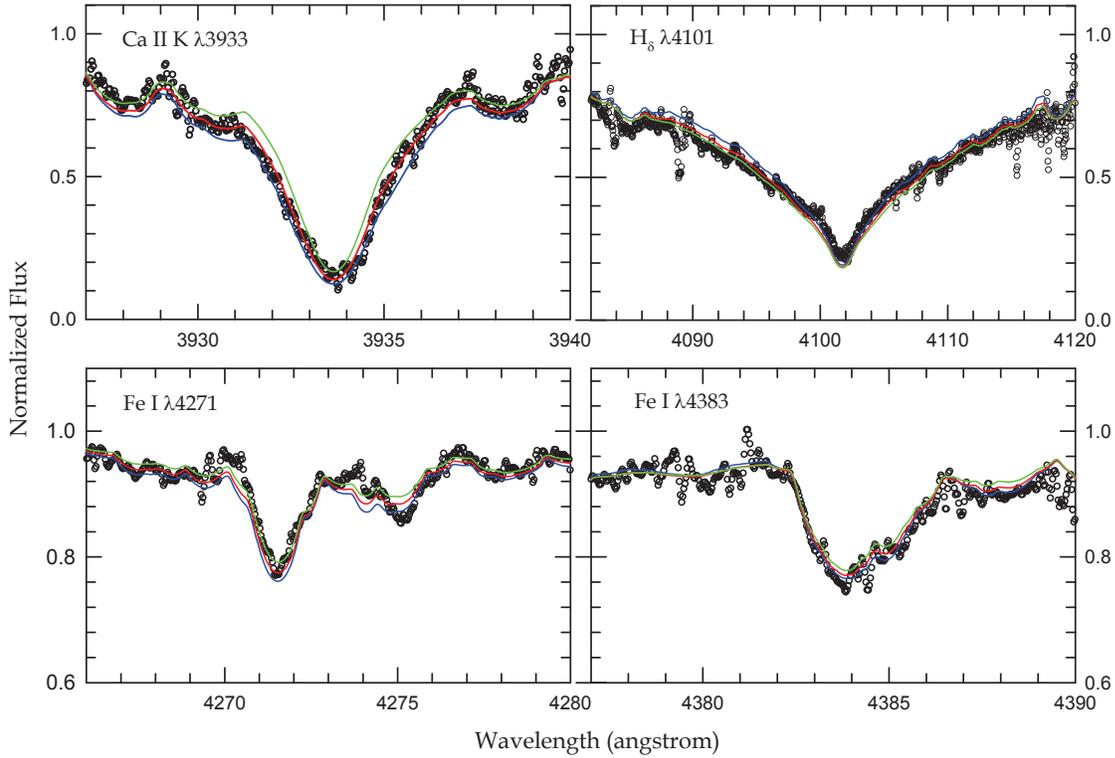}
\caption{Four spectral regions of the primary star. The open circles represent the disentangling spectrum obtained by 
the FDBinary code. The blue, red, and green lines represent the synthetic spectra of 8050 K, 8260 K, and 8470 K, respectively, 
interpolated from the atmosphere models of Kurucz (1993), where log $g_1 = 4.1$ and [Fe/H] = 0.0. }
\label{Fig4}
\end{figure}

\begin{figure}
\includegraphics[]{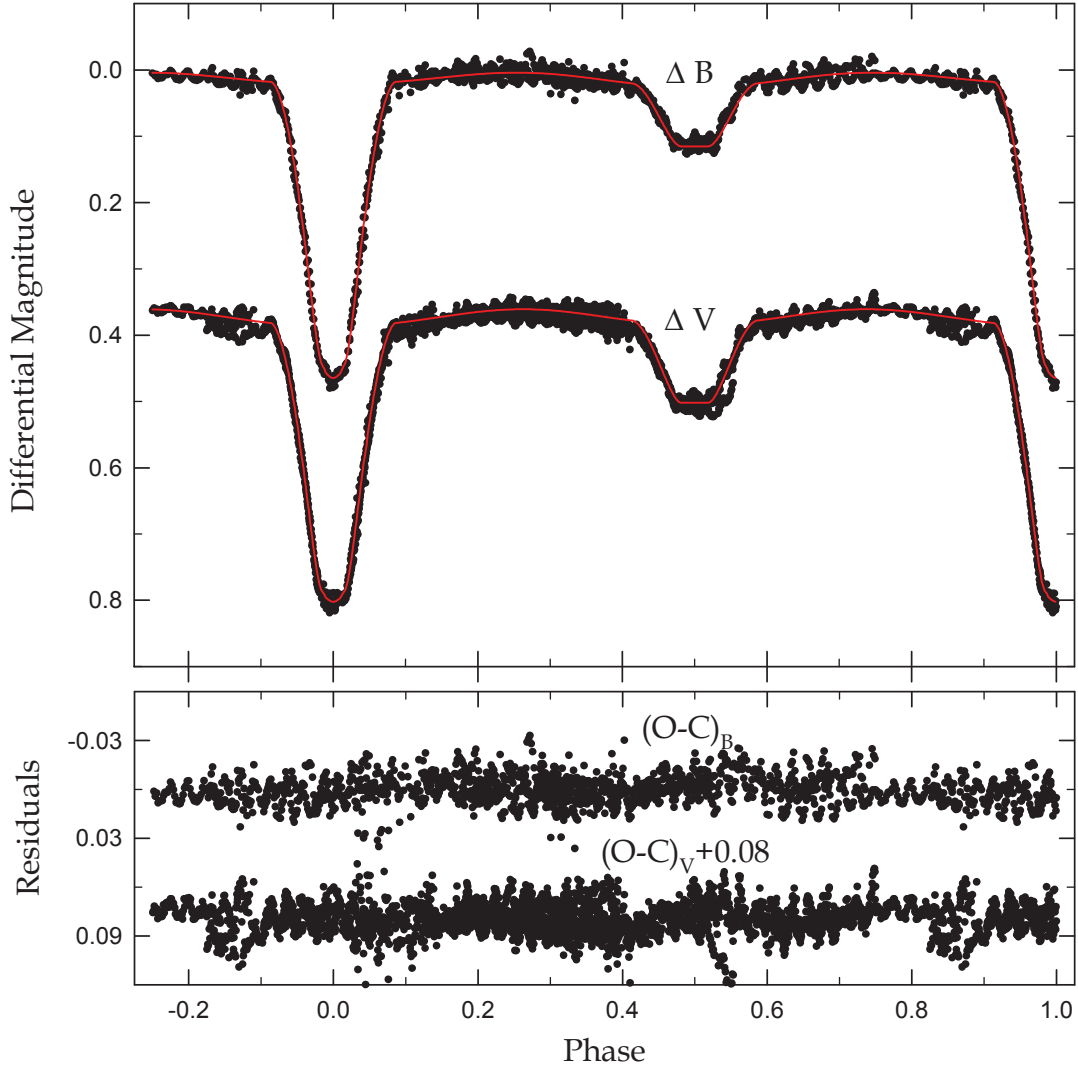}
\caption{$BV$ light curves of OO Dra with fitted models. The circles are individual measures taken from Zhang et al. (2014), 
and the solid lines represent the synthetic curves obtained from a simultaneous analysis of the light and RV curves. 
The lower panel shows the differences between observations and theoretical models. }
\label{Fig5}
\end{figure}

\begin{figure}
\includegraphics[scale=0.9]{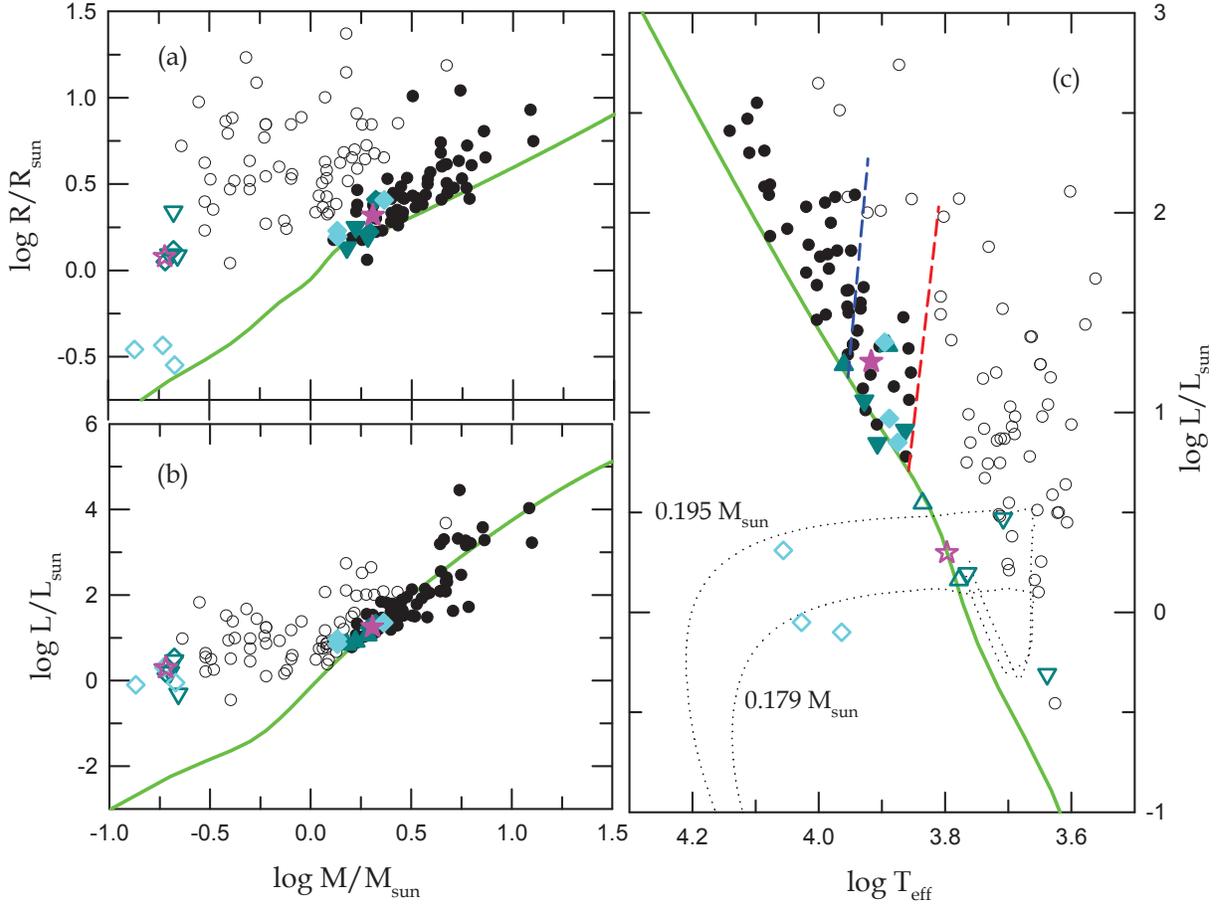}
\caption{Plots of (a) mass-radius, (b) mass-luminosity, and (c) HR diagrams for OO Dra (star symbols), semi-detached Algols 
(circles), and R CMa-type (triangles) and EL CVn-type (diamonds) EBs. The filled and open symbols represent the primary and 
secondary components, respectively. The solid green lines denote the ZAMS for solar metallicity from Tout et al. (1996). 
In panel (c), the dashed blue and red lines represent the theoretical edges of the $\delta$ Sct instability strip 
(Rolland et al. 2002; Soydugan et al. 2006), and the dotted black lines represent the evolutionary track for He WDs with masses of 
0.179 $M_\odot$ and 0.195 $M_\odot$ (Driebe et al. 1998). }
\label{Fig6}
\end{figure}

\clearpage                                                                                                           
\begin{deluxetable}{lcccc}
\tablewidth{0pt}                    
\tabletypesize{\scriptsize}                                                                                  
\tablecaption{Radial Velocities of OO Dra. }                                                                            
\tablehead{    
\colhead{HJD}          & \colhead{$V_{1}$}       & \colhead{$\sigma_1$}    &  \colhead{$V_{2}$}      &  \colhead{$\sigma_2$}    \\                                            
\colhead{(2,457,000+)} & \colhead{(km s$^{-1}$)} & \colhead{(km s$^{-1}$)} & \colhead{(km s$^{-1}$)} & \colhead{(km s$^{-1}$)}  
}                                                                                                 
\startdata
442.0773               &  $ -38.3 $             &  21.2                  &  $ \dots   $            &  \dots                   \\
442.0984               &  $ -42.6 $             &  11.6                  &  $ \dots   $            &  \dots                   \\
442.9499               &  $ -38.2 $             &  7.2                   &  $ -227.3  $            &  7.3                     \\
443.9326               &  $ -64.4 $             &  7.2                   &  $ 63.6    $            &  9.2                     \\
443.9537               &  $ -62.5 $             &  5.3                   &  $ \dots   $            &  \dots                   \\
443.9748               &  $ -57.5 $             &  3.0                   &  $ 36.2    $            &  12.3                    \\
443.9959               &  $ -55.5 $             &  4.0                   &  $ \dots   $            &  \dots                   \\
444.0170               &  $ -47.3 $             &  9.6                   &  $ \dots   $            &  \dots                   \\
444.0382               &  $ -53.7 $             &  13.5                  &  $ \dots   $            &  \dots                   \\
444.0593               &  $ -44.2 $             &  9.0                   &  $ \dots   $            &  \dots                   \\
444.0804               &  $ -52.5 $             &  2.6                   &  $ \dots   $            &  \dots                   \\
496.9744               &  $ -77.2 $             &  6.2                   &  $ 191.9   $            &  13.0                    \\
496.9955               &  $ -76.6 $             &  8.5                   &  $ 186.4   $            &  18.2                    \\
497.1776               &  $ -60.8 $             &  6.7                   &  $ 64.5    $            &  12.3                    \\
498.0150               &  $ -64.4 $             &  9.0                   &  $ 48.0    $            &  14.6                    \\
498.0370               &  $ -69.0 $             &  7.4                   &  $ 83.2    $            &  7.5                     \\
498.0652               &  $ -67.6 $             &  6.3                   &  $ 104.4   $            &  17.8                    \\
498.0864               &  $ -70.0 $             &  3.0                   &  $ 132.3   $            &  8.4                     \\
498.1075               &  $ -72.1 $             &  6.6                   &  $ 149.0   $            &  18.5                    \\
498.1287               &  $ -69.0 $             &  7.6                   &  $ 148.4   $            &  7.6                     \\
498.1506               &  $ -72.9 $             &  9.8                   &  $ 160.8   $            &  8.7                     \\
498.1718               &  $ -75.4 $             &  4.7                   &  $ 157.5   $            &  4.7                     \\
498.1929               &  $ -71.8 $             &  6.9                   &  $ 182.8   $            &  12.0                    \\
498.2141               &  $ -71.4 $             &  4.9                   &  $ 179.9   $            &  9.7                     \\
498.2352               &  $ -75.4 $             &  12.7                  &  $ 181.0   $            &  12.6                    \\
498.2564               &  $ -72.3 $             &  7.2                   &  $ 179.5   $            &  11.7                    \\
498.2779               &  $ -76.8 $             &  4.9                   &  $ 173.2   $            &  9.8                     \\
498.2990               &  $ -74.4 $             &  12.2                  &  $ \dots   $            &  \dots                   \\
498.3202               &  $ -64.5 $             &  14.8                  &  $ \dots   $            &  \dots                   \\
522.0028               &  $ -56.8 $             &  5.5                   &  $ \dots   $            &  \dots                   \\
522.0240               &  $ -58.7 $             &  5.8                   &  $ \dots   $            &  \dots                   \\
522.0451               &  $ -59.5 $             &  6.7                   &  $ \dots   $            &  \dots                   \\
522.0663               &  $ -54.7 $             &  4.2                   &  $ \dots   $            &  \dots                   \\
522.0874               &  $ -53.8 $             &  4.0                   &  $ \dots   $            &  \dots                   \\
522.1086               &  $ -54.9 $             &  5.3                   &  $ -110.4  $            &  7.1                     \\
526.0577               &  $ -30.3 $             &  4.4                   &  $ -283.5  $            &  9.0                     \\
526.0818               &  $ -30.8 $             &  3.1                   &  $ -290.9  $            &  7.6                     \\
526.1083               &  $ -29.0 $             &  4.7                   &  $ -285.6  $            &  14.2                    \\
526.1296               &  $ -31.8 $             &  7.4                   &  $ -276.5  $            &  18.6                    \\
\enddata                                                                                                             
\end{deluxetable}

\begin{deluxetable}{lcc}
\tablewidth{0pt}
\tablecaption{Orbital Elements of OO Dra Derived with Sine-Curve Fits. }
\tablehead{
\colhead{Parameter}          & \colhead{Primary}         & \colhead{Secondary}
}                                                         
\startdata                                                
$T_0$ (HJD)                  & \multicolumn{2}{c}{2,457,444.6619$\pm$0.0018}         \\
$P$ (d)$\rm ^a$              & \multicolumn{2}{c}{1.238378}                          \\
$\gamma$ (km s$^{-1}$)       & $-$52.5$\pm$1.6           & $-$55.0$\pm$3.8           \\
$K$ (km s$^{-1}$)            & 21.7$\pm$2.3              & 235.1$\pm$4.5             \\ 
$a\sin$$i$ ($R_\odot$)       & 0.53$\pm$0.06             & 5.75$\pm$0.11             \\
$M\sin ^3$$i$ ($M_\odot$)    & 1.989$\pm$0.087           & 0.184$\pm$0.024           \\
$q $ (= $M_2/M_1$)           & \multicolumn{2}{c}{0.092$\pm$0.010}                   \\
\enddata
\tablenotetext{a}{Fixed.}
\end{deluxetable}

\begin{deluxetable}{lccc}
\tablewidth{0pt}
\tablecaption{Light-Curve and RV Sets for OO Dra.}
\tablehead{
\colhead{Reference}          & \colhead{Season} & \colhead{Data type} & \colhead{$\sigma$} }
\startdata
Zhang et al. (2014)          & 2014             & $B$                 & 0.0090 mag         \\
                             &                  & $V$                 & 0.0092 mag         \\
This paper                   & 2016             & RV1                 & 3.3 km s$^{-1}$    \\
                             &                  & RV2                 & 7.8 km s$^{-1}$    \\
\enddata
\end{deluxetable}

\begin{deluxetable}{lcc}
\tablewidth{0pt} 
\tablecaption{Light and RV Parameters of OO Dra}
\tablehead{
\colhead{Parameter}                      & \colhead{Primary} & \colhead{Secondary}                                                  
}                                                                                                                                     
\startdata                                                                                                                            
$T_0$ (BJD)                              & \multicolumn{2}{c}{2,456,741.26035$\pm$0.00011}   \\
$P$ (day)                                & \multicolumn{2}{c}{1.2383813$\pm$0.0000050}       \\
$a$ (R$_\odot$)                          & \multicolumn{2}{c}{6.326$\pm$0.063}               \\
$\gamma$ (km s$^{-1}$)                   & \multicolumn{2}{c}{$-$52.84$\pm$0.57}             \\
$K_1$ (km s$^{-1}$)                      & \multicolumn{2}{c}{21.7$\pm$1.0}                  \\
$K_2$ (km s$^{-1}$)                      & \multicolumn{2}{c}{236.0$\pm$3.1}                 \\
$q$                                      & \multicolumn{2}{c}{0.0919$\pm$0.0034}             \\
$i$ (deg)                                & \multicolumn{2}{c}{85.320$\pm$0.075}              \\
$T$ (K)                                  & 8260$\pm$210       & 6268$\pm$150                 \\
$\Omega$                                 & 3.1744$\pm$0.0041  & 1.9564$\pm$0.0010            \\
$\Omega_{\rm in}$                        & \multicolumn{2}{c}{1.9339}                        \\
$A$                                      & 1.0                & 0.5                          \\
$g$                                      & 1.0                & 0.32                         \\
$X$, $Y$                                 & 0.659, 0.144       & 0.636, 0.241                 \\
$x_{B}$, $y_{B}$                         & 0.801, 0.317       & 0.812, 0.195                 \\
$x_{V}$, $y_{V}$                         & 0.694, 0.286       & 0.721, 0.263                 \\
$L/(L_1+L_2)_{B}$                        & 0.9294$\pm$0.0004  & 0.0706                       \\
$L/(L_1+L_2)_{V}$                        & 0.9027$\pm$0.0004  & 0.0973                       \\
$r$ (pole)                               & 0.3239$\pm$0.0004  & 0.1776$\pm$0.0007            \\
$r$ (point)                              & 0.3326$\pm$0.0005  & 0.2242$\pm$0.0025            \\
$r$ (side)                               & 0.3303$\pm$0.0005  & 0.1837$\pm$0.0008            \\
$r$ (back)                               & 0.3317$\pm$0.0005  & 0.2074$\pm$0.0014            \\
$r$ (volume)$\rm ^a$                     & 0.3287$\pm$0.0005  & 0.1896$\pm$0.0012            \\ 
\enddata
\tablenotetext{a}{Mean volume radius.}
\end{deluxetable}

\begin{deluxetable}{lcccccccc}
\tablewidth{0pt} 
\tablecaption{Absolute Parameters of OO Dra}
\tablehead{
\colhead{Parameter}                      & \colhead{Primary} & \colhead{Secondary}                                                  
}                                                                                                                                     
\startdata                                                                                                                            
$M$ ($M_\odot$)                          & 2.031$\pm$0.058    & 0.187$\pm$0.009              \\
$R$ ($R_\odot$)                          & 2.078$\pm$0.026    & 1.199$\pm$0.017              \\
$\log$ $g$ (cgs)                         & 4.110$\pm$0.017    & 3.552$\pm$0.026              \\
$\rho$ (g cm$^3)$                        & 0.320$\pm$0.015    & 0.153$\pm$0.010              \\
$L$ ($L_\odot$)                          & 18.0$\pm$1.9       & 1.99$\pm$0.20                \\
$M_{\rm bol}$ (mag)                      & 1.59$\pm$0.11      & 3.99$\pm$0.11                \\
BC (mag)                                 & 0.02$\pm$0.01      & $-$0.01$\pm$0.01             \\
$M_{\rm V}$ (mag)                        & 1.57$\pm$0.11      & 4.00$\pm$0.11                \\
Distance (pc)                            & \multicolumn{2}{c}{760$\pm$42}                    \\
\enddata
\end{deluxetable}


\newpage
\begin{thebibliography}{}
\bibitem[Biyalieva \& Khruslov(2007)]{biyalieva2008} Biyalieva, N. A., \& Khruslov, A. V. 2007, PZP, 7, 17
\bibitem[Breger(2000)]{breger1993} Breger, M. 2000, in ASP Conf. Ser. 210, Delta Scuti and Related Stars, ed. M. Breger, \& M. H. Montgomery (San Francisco, CA: ASP), 3
\bibitem[Budding \& Butland(2011)]{budding2011} Budding, E., \& Butland, R. 2011, MNRAS, 418, 1764
\bibitem[Cakirli \& Ibanoglu(2017)]{cakirli2016} \c Cakirli, \" O., \& Ibano\v{g}lu, C. 2016, New Astron., 45, 36
\bibitem[Chen et al(2017)]{chen2017} Chen, X., Maxted, P. F. L., Li, J., \& Han, Z. 2017, MNRAS, 467, 1874
\bibitem[Dimitrov et al(2008)]{dimitrov2008} Dimitrov, D., Kraicheva, Z., \& Popov, V. 2008, IBVS, 5842, 1
\bibitem[Driebe et al(1998)]{driebe1998} Driebe, T., Sch\"oenberner, D., Bl\"oecker, T., \& Herwig, F., 1998, A\&A, 339, 123
\bibitem[Erdem \& Ozturk(2014)]{erdem2014} Erdem, A., \& \"Ozt\"urk, O. 2014, MNRAS, 441, 1166
\bibitem[Fitch(1981)]{fitch1981} Fitch, W. S. 1981, ApJ, 249, 218
\bibitem[GAIA(2016)]{gaia2016} Gaia Collaboration, Brown, A. G. A., Vallenari, A., et al. 2016, A\&A, 595, 2
\bibitem[Gaulme(2013)]{gaulme2013} Gaulme, P., McKeever, J., Rawls, M. L., et al. 2013, ApJ, 767, 82
\bibitem[Guo et al(2017)]{guo2017} Guo, Z., Gies, D. R., Matson, R. A., Garcia Hernandez, A., Han, Z., \& Chen, X. 2017, ApJ, 837, 114
\bibitem[Henden et al(2016)]{henden2016} Henden, A., Templeton, M., Terrell, D., et al. 2016, yCat, 2336, 0
\bibitem[Hilditch(2001)]{hilditch2001} Hilditch, R. W. 2001, An Introduction to Close Binary Stars (Cambridge: Cambridge Univ. Press)
\bibitem[Hong et al(2015)]{hong2015} Hong, K., Lee, J. W., Kim, S.-L., et al. 2015, AJ, 150, 131
\bibitem[Hong et al(2017)]{hong2017} Hong, K., Lee, J. W., Koo, J.-R., et al. 2017, AJ, 153, 247
\bibitem[Ibanoglu et al(2006)]{ibanoglu2006} \. Ibano\v{g}lu, C., Soydugan, F., Soydugan, E., \& Dervi\c so\v{g}lu, A. 2006, MNRAS, 373, 435
\bibitem[Ilijic et al(2004)]{ilijic2004} Iliji\'c, S., Hensberge, H., Pavlovski, K., \& Freyhammer, L. M. 2004, in ASP Conf. Ser. 318, Spectroscopically and Spatially Resolving the Components of the Close Binary Stars, ed. R. Hilditch, H. Hensberge, \& K. Pavlovski (San Francisco, CA: ASP), 111 
\bibitem[Kahraman Alicavus et al(2017)]{kahraman2017} Kahraman Ali\c cavu\c s, F., Soydugan, E., Smalley, B., \& Kub\'at, J. 2017, MNRAS, 470, 915 
\bibitem[Kaluzny et al(2007)]{kaluzny2007} Kaluzny, J., Thompson, I. B., Rucinski, S. M., et al. 2007, AJ, 134, 541
\bibitem[Kim et al(2007)]{kim2007} Kim, K.-M., Han, I., Valyavin, G. G., et al. 2007, PASP, 119, 1052
\bibitem[Koo et al(2016)]{koo2016} Koo, J.-R., Lee, J. W., Hong, K., Kim, S.-L., \& Lee, C.-U. 2016, AJ, 151, 77
\bibitem[Kurucz(1993)]{kurucz1993} Kurucz, R. L. 1993, CD-ROMs No. 1-18 (Cambridge, MA: Smithsonian Astrophysical Observatory) 
\bibitem[Lee et al(2008)]{lee2008} Lee, C.-U., Kim, S.-L., Lee, J. W., et al. 2008, MNRAS, 389, 1630
\bibitem[Lee(2016)]{lee2016a} Lee, J. W. 2016, ApJ, 833, 170
\bibitem[Lee et al(2016)]{lee2016b} Lee, J. W., Hong, K., Kim, S.-L., \& Koo, J.-R. 2016, MNRAS, 460, 4220
\bibitem[Lee et al(2014)]{lee2014} Lee, J. W., Park, J.-H., Hong, K., Kim, S.-L., \& Lee, C.-U., 2014, AJ, 147, 91
\bibitem[Lee et al(2015)]{lee2015} Lee, J. W., Youn, J.-H., Park, J.-H., \& Wolf, M. 2015, AJ, 149, 194
\bibitem[Lehmann et al(2013)]{lehmann2013} Lehmann, H., Southworth, J., Tkachenko, A., \& Pavlovski, K. 2013, A\&A, 557, A79
\bibitem[Liakos \& Niarchos(2017)]{liakos2017} Liakos, A., \& Niarchos, P. 2017, MNRAS, 465, 1181
\bibitem[Lucy(1967)]{lucy1967} Lucy, L. B. 1967, Z. Astrophys., 65, 89
\bibitem[Matson et al(2015)]{matson2015} Matson, R. A., Gies, D. R., Guo, Z., et al. 2015, ApJ, 806, 155 
\bibitem[Maxted et al(2014a)]{maxted2014a} Maxted, P. F. L., Bloemen, S., Heber, U., et al. 2014a, MNRAS, 437, 168
\bibitem[Maxted et al(2014b)]{maxted2014a} Maxted, P. F. L., Serenelli, A. M., Marsh, T. R., Catal\'an, S., Mahtani, D. P., \& Dhillon, V. S., 2014b, MNRAS, 444, 208
\bibitem[Maxted et al(2013)]{maxted2013} Maxted, P. F. L., Serenelli, A. M., Miglio, A., et al. 2013, Natur, 498, 463
\bibitem[Mkrtichian et al(2004)]{mkrtichian2004} Mkrtichian, D. E., Kusakin, A. V., Rodriguez, E., et al. 2004, A\&A, 419, 1015
\bibitem[Petersen \& Jorgensen(1972)]{petersen1972} Petersen, J. O., \& J\o rgensen, H. E. 1972, A\&A, 17, 367
\bibitem[Rolland et al(2002)]{rolland2002} Rolland, A., Costa, V., Rodr\'iguez, E., et al. 2002, Comm. Asteroseismology, 142, 57
\bibitem[Rucinski(1969a)]{rucinski1969a} Rucinski, S. M. 1969a, Acta. Astron., 19, 125
\bibitem[Rucinski(1969b)]{rucinski1969b} Rucinski, S. M. 1969b, Acta. Astron., 19, 245
\bibitem[Sarna(1993)]{sarna1993} Sarna, M. J. 1993, MNRAS, 262, 534 
\bibitem[Schaffenroth et al(2015)]{schaffenroth2015} Schaffenroth, V., Barlow, B. N., Drechsel, H., \& Dunlap, B. H. 2015, A\&A, 576, 123
\bibitem[Schlafly \& Finkbeiner(2011)]{schlafly2011} Schlafly, E. F., \& Finkbeiner, D. P. 2011, AJ, 737, 103
\bibitem[Soydugan et al(2006)]{soydugan2006} Soydugan, E., \. Ibano\v{g}lu, C., Soydugan, F., Akan, M. C., \& Demircan, O. 2006, MNRAS, 366, 1298
\bibitem[Torres(2010)]{torres2010} Torres, G. 2010, AJ, 140, 1158
\bibitem[Tout(1996)]{tout1996} Tout, C. A., Pols, O. R., Eggleton, P. P., \& Han, Z. 1996, MNRAS, 281, 257
\bibitem[Van Hamme(1993)]{van1993} Van Hamme, W. 1993, AJ, 106, 209
\bibitem[Van Hamme \& Wilson(1984)]{van1984} Van Hamme, W., \& Wilson, R. E. 1984, A\&A, 141, 1 
\bibitem[Van Hamme \& Wilson(2007)]{van2007} Van Hamme, W., \& Wilson, R. E. 2007, ApJ, 661, 1129
\bibitem[Von Zeipel(1924)]{von1924} Von Zeipel, H., 1924, MNRAS, 84, 665
\bibitem[Wilson \& Devinney(1971)]{wilson1971} Wilson, R. E., \& Devinney, E. J. 1971, ApJ, 166, 605
\bibitem[Zhang et al(2014)]{Zhang2014} Zhang, X. B., Deng, L, C., Tian, J. F., et al. 2014, AJ, 148, 106
\bibitem[Zhang et al(2016)]{Zhang2016} Zhang, X. B., Fu, J. N., Li, Y., Ren, A. B., \& Luo, C. Q. 2016, ApJL, 821, L32
\bibitem[Zhang et al(2013)]{Zhang2013} Zhang, X. B., Luo, C. Q., \& Fu, J. N. 2013, ApJ, 777, 77
\bibitem[Zhou(2010)]{Zhou2010} Zhou, A. Y. 2010, arXiv:1002.2729v6 
\end{thebibliography}
\end{document}